\documentclass[twoside,twocolumn]{article}
\makeatletter\if@twocolumn\PassOptionsToPackage{switch}{lineno}\else\fi\makeatother

\usepackage{amsfonts,amssymb,amsbsy,latexsym,amsmath,tabulary,graphicx,times,xcolor}

\usepackage[nonindentfirst]{titlesec}
\usepackage{caption,abstract,fancyhdr}
\usepackage[utf8]{inputenc}
\usepackage{balance}

\usepackage{colortbl}
  \newcommand{\myrowcolour}{\rowcolor[gray]{0.925}}

\def\oupIndent{1pt}

\usepackage[T1]{fontenc}
\usepackage[paperheight=11.69in,paperwidth=8.27in,margin=1.5cm,headsep=.5cm,top=2cm,bottom=2.5cm,columnsep=12pt]{geometry}
 \date{}
\def\abstractname{\fontsize{16pt}{19.2pt}\sffamily{Abstract}\selectfont}
\renewenvironment{onecolabstract} {\vspace*{-1pc}\trivlist\item[]\leftskip\oupIndent\par\vskip4pt\noindent\textit{{\abstractname}}\mbox{\null}\\ \textcolor[cmyk]{.77,.49,.38,.11}{\rule[.5pc]{\textwidth}{2pt}} \newline}{\par\noindent\endtrivlist}

\makeatletter
\def\author#1{\gdef\@author{\hskip-\dimexpr(\tabcolsep)\hskip\oupIndent\parbox{\dimexpr\textwidth-\oupIndent}{#1}}}
\def\title#1{\gdef\@title{\raggedright\sffamily\bfseries\ifx\@articleType\@empty\else\@articleType\\\fi#1}}
\fancypagestyle{headings}{
\fancyhf{}\fancyhead[R]{{\rule{\textwidth}{.1pt}} \\ \RunningAuthor }\fancyhead[L]{{\rule{\textwidth}{0pt}} \\ \MakeUppercase{\journalTitle} }\fancyfoot[R]{\thepage}}
\fancypagestyle{plain}{\fancyhf{}\fancyhead[R]{{\rule{\textwidth}{.1pt}} \\ \RunningAuthor }\fancyhead[L]{{\rule{\textwidth}{0pt}} \\ \MakeUppercase{\journalTitle}}\fancyfoot[R]{\thepage}}
\let\@articleType\@empty \def\articletype#1{\gdef\@articleType{{\normalfont\underline{#1}}}}

\usepackage{titlesec}
 \def\NormalBaseline{\def\baselinestretch{1.1}}
\titleformat{\section}[hang]{\NormalBaseline\filright\large\fontsize{16pt}{19.2pt}\itshape\sffamily}
{\thesection}
{10pt}
{}
[{\color[cmyk]{.77,.49,.38,.11}\rule[3pc]{\columnwidth}{2pt}}]
\titleformat{\subsection}[hang]{\NormalBaseline\bfseries}
{\thesubsection}
{10pt}
{}
[]
\titleformat{\subsubsection}[hang]{\NormalBaseline\bfseries}
{\thesubsubsection}
{10pt}
{}
[]
\titleformat{\paragraph}[runin]{\NormalBaseline\bfseries}
{\theparagraph}
{10pt}
{}
[]
\titleformat{\subparagraph}[runin]{\NormalBaseline\filright\bfseries}
{\thesubparagraph}
{10pt}
{}
[]

\titlespacing{\section}{0pt}{1.5\baselineskip}{-2.5\baselineskip}
\titlespacing{\subsection}{0pt}{1.5\baselineskip}{.2\baselineskip}
\titlespacing{\subsubsection}{0pt}{1.5\baselineskip}{.2\baselineskip}
\titlespacing{\paragraph}{0pt}{.5\baselineskip}{10pt}
\titlespacing{\subparagraph}{0pt}{.5\baselineskip}{10pt}
\makeatother

\usepackage{url,multirow,morefloats,floatflt,cancel,tfrupee}
\makeatletter

\AtBeginDocument{\@ifpackageloaded{textcomp}{}{\usepackage{textcomp}}}
\makeatother
\usepackage{colortbl}
\usepackage{xcolor}
\usepackage{pifont}
\usepackage[nointegrals]{wasysym}
\makeatletter

\def\mcWidth#1{\csname TY@F#1\endcsname+\tabcolsep}

\def\cAlignHack{\rightskip\@flushglue\leftskip\@flushglue\parindent\z@\parfillskip\z@skip}
\def\rAlignHack{\rightskip\z@skip\leftskip\@flushglue \parindent\z@\parfillskip\z@skip}

\@ifundefined{etal}{}{}

\usepackage{ifxetex}
\ifxetex\else\if@twocolumn\@ifpackageloaded{stfloats}{}{\usepackage{dblfloatfix}}\fi\fi

\AtBeginDocument{
\expandafter\ifx\csname eqalign\endcsname\relax
\def\eqalign#1{\null\vcenter{\def\\{\cr}\openup\jot\m@th
  \ialign{\strut$\displaystyle{##}$\hfil&$\displaystyle{{}##}$\hfil
      \crcr#1\crcr}}\,}
\fi
}

\AtBeginDocument{%
  \@ifpackageloaded{endfloat}%
   {\renewcommand\efloat@iwrite[1]{\immediate\expandafter\protected@write\csname efloat@post#1\endcsname{}}}{\newif\ifefloat@tables}%
}%

\def\BreakURLText#1{\@tfor\brk@tempa:=#1\do{\brk@tempa\hskip0pt}}
\let\lt=<
\let\gt=>
\def\processVert{\ifmmode|\else\textbar\fi}

\@ifundefined{subparagraph}{
\def\subparagraph{\@startsection{paragraph}{5}{2\parindent}{0ex plus 0.1ex minus 0.1ex}%
{0ex}{\normalfont\small\itshape}}%
}{}

\newcommand\role[1]{\unskip}
\newcommand\aucollab[1]{\unskip}
  
\@ifundefined{tsGraphicsScaleX}{\gdef\tsGraphicsScaleX{1}}{}
\@ifundefined{tsGraphicsScaleY}{\gdef\tsGraphicsScaleY{.9}}{}
\def\checkGraphicsWidth{\ifdim\Gin@nat@width>\linewidth
	\tsGraphicsScaleX\linewidth\else\Gin@nat@width\fi}

\def\checkGraphicsHeight{\ifdim\Gin@nat@height>.9\textheight
	\tsGraphicsScaleY\textheight\else\Gin@nat@height\fi}

\def\fixFloatSize#1{}
\let\ts@includegraphics\includegraphics

\def\inlinegraphic[#1]#2{{\edef\@tempa{#1}\edef\baseline@shift{\ifx\@tempa\@empty0\else#1\fi}\edef\tempZ{\the\numexpr(\numexpr(\baseline@shift*\f@size/100))}\protect\raisebox{\tempZ pt}{\ts@includegraphics{#2}}}}

\AtBeginDocument{\def\includegraphics{\@ifnextchar[{\ts@includegraphics}{\ts@includegraphics[width=\checkGraphicsWidth,height=\checkGraphicsHeight,keepaspectratio]}}}

\DeclareMathAlphabet{\mathpzc}{OT1}{pzc}{m}{it}

\def\URL#1#2{\@ifundefined{href}{#2}{\href{#1}{#2}}}

\def\UrlOrds{\do\*\do\-\do\~\do\'\do\"\do\-}%
\g@addto@macro{\UrlBreaks}{\UrlOrds}

\edef\fntEncoding{\f@encoding}

\makeatother

\newif\ifmultipleabstract\multipleabstractfalse%
\usepackage[numbers,sort&compress]{natbib}

\begin{document}

\title{Predicting Patient COVID-19 Disease Severity by means of Statistical and Machine Learning Analysis of Blood Cell Transcriptome Data}
\author{\textcolor{black}{\rule{\textwidth}{.1pt}}\vspace*{-.5pc} \textcolor{black}{\rule{\textwidth}{.1pt}} Sakifa Aktar \textsuperscript{1}, B.Sc.;
            Md. Martuza Ahamad \textsuperscript{1}, M.Sc.;
            Md. Rashed-Al-Mahfuz \textsuperscript{2}, M.Sc.;
            AKM Azad \textsuperscript{3}, PhD;
            Shahadat Uddin \textsuperscript{4}, PhD;
            A H M Kamal \textsuperscript{5}, Prof.;
            Salem A. Alyami \textsuperscript{6}, PhD;
            Ping-I Lin \textsuperscript{7}, PhD;
            Sheikh Mohammed Shariful Islam \textsuperscript{8}, PhD;
            Julian M.W. Quinn \textsuperscript{9}, PhD;
            Valsamma Eapen \textsuperscript{7,*}, Prof.;
            Mohammad Ali Moni \textsuperscript{7,9,10,*}, PhD\vspace*{-.5pc}\\\vspace*{-1pc}\textcolor{black}{\rule{\textwidth}{.1pt}}~\\[-3pt]\normalsize 
~\\\textsuperscript{1}{Department of Computer Science and Engineering\unskip, Bangabandhu Sheikh Mujibur Rahman Science \& Technology University\unskip, Gopalgamj\unskip, 8100\unskip, Bangladesh}
~\\\textsuperscript{2}{Department of Computer Science and Engineering\unskip, University of Rajshahi\unskip, Rajshahi\unskip, 6205\unskip, Bangladesh}, Email: ram@ru.ac.bd
~\\\textsuperscript{3}{iThree Institute, Faculty of Science\unskip, University of Technology Sydney\unskip, Australia}
~\\\textsuperscript{4}{Complex Systems Research Group, Faculty of Engineering\unskip, The University of Sydney, Darlington, NSW\unskip, 2008\unskip, Australia}
~\\\textsuperscript{5}{Department of Computer Science and Engineering\unskip, Jatiya Kabi Kazi Nazrul Islam University\unskip, Mymensingh\unskip, Bangladesh}
~\\\textsuperscript{6}{Department of Mathematics and Statistics\unskip, Imam Mohammad Ibn Saud Islamic University\unskip, Saudi Arabia}
~\\\textsuperscript{7}{Faculty of Medicine, School of Psychiatry\unskip, University of New South Wales, Sydney, NSW\unskip, Australia}
~\\\textsuperscript{8}{Institute for Physical Activity and Nutrition (IPAN), Faculty of Health\unskip, Deakin University}
~\\\textsuperscript{9}{Healthy Ageing Theme, Darlinghurst, NSW\unskip, The Garvan Institute of Medical Research\unskip, Australia}
~\\\textsuperscript{10}{WHO Collaborating Centre on eHealth, UNSW Digital Health, School of Public Health and Community Medicine, Faculty of Medicine, UNSW Sydney\unskip, Australia}}

\def\RunningAuthor{Sakifa Aktar et~al.}\def\journalTitle{}

\twocolumn[ \maketitle {\begin{onecolabstract}
 \textbf{Introduction:} For COVID-19 patients' accurate prediction of disease severity and mortality risk would greatly improve care delivery and resource allocation. There are many patient-related factors, such as pre-existing comorbidities that affect disease severity. Since rapid automated profiling of peripheral blood samples is widely available, we investigated how such data from the peripheral blood of COVID-19 patients might be used to predict clinical outcomes.

\textbf{Methods:} We thus investigated such clinical datasets from COVID-19 patients with known outcomes by combining statistical comparison and correlation methods with machine learning algorithms; the latter included decision tree, random forest, variants of gradient boosting machine, support vector machine, K-nearest neighbour and deep learning methods.

\textbf{Results:} Our work revealed several clinical parameters measurable in blood samples, which discriminated between healthy people and COVID-19 positive patients and showed predictive value for later severity of COVID-19 symptoms. We thus developed a number of analytic methods that showed accuracy and precision for disease severity and mortality outcome predictions that were above 90\%.

\textbf{Conclusions:} In sum, we developed methodologies to analyse patient routine clinical data which enables more accurate prediction of COVID-19 patient outcomes. This type of approaches could, by employing standard hospital laboratory analyses of patient blood, be utilised to identify, COVID-19 patients at high risk of mortality and so enable their treatment to be optimised.

\def\keywordstitle{Keywords}
\textcolor{black}{\rule{\textwidth}{.2pt}} \\ \smallskip\noindent\textbf{KEYWORDS: }{COVID-19; Blood Samples; Machine learning; Statistical analysis\newline}
\smallskip\noindent\textbf{*Corresponding Authors: }{Mohammad Ali Moni (m.moni@unsw.edu.au),  Valsama Eapen (v.eapen@unsw.edu.au)\newline}
\end{onecolabstract}}]\saythanks 
    
\section*{Introduction}
 Severe acute respiratory syndrome coronavirus-2 (SARS-CoV-2) has caused the current pandemic of coronavirus disease-19 (COVID-19), which first emerged as an outbreak in December 2019 in the Chinese province of Hubei [1]. The management of COVID-19 infected patients remains problematic and controversial, although this is to be expected in such a recently emerged disease. The first symptoms of COVID-19 resemble those of many other infections and inflammatory conditions that affect the respiratory system and include fever, sneezing and rhinitis, persistent cough and fatigue with general aches and pains [2]. However, an infected patient can rapidly develop further more severe symptoms which can be life-threatening and require intensive care intervention; these include pneumonia, severe shortage of breath, diarrhoea, dispersed thromboses and vascular inflammation [3]. A further issue in caring for COVID-19 patients is the presence of comorbidities that interact with COVID-19, particularly pulmonary and vascular conditions, that can make the patient prognosis far worse. This is an important consideration given the current lack of availability of a vaccine or effective therapy for COVID-19. However, there have been notable advances in treating patients with advanced disease and indeed the common recognition of the importance that the primary defenses against COVID-19 are public health measures have reduced its spread.

Intensive care units (ICUs) are key to improving the survival of patients with severe COVID-19, providing oxygen, 24-hour monitoring and care and assisted ventilation when needed. Therefore thus, ICU beds are a precious resource in locations where COVID-19 case numbers are high [4]. Therefore, allocating hospital wards or ICU beds for infected patients require rapid decision-making processes, both to utilise resources efficiently and reduce patient suffering and mortality. In many parts of the world, stressed care systems face significant difficulty in deciding on ICU bed allocation, so a smart, automated system could be useful to improve care and resource allocation. The World Health Organization has recommended that all suspected COVID-19 patients be tested by reverse transcription-polymerase chain reaction (RT-PCR) based diagnosis methods that directly detect viral RNA [5]. Testing by approaches other than RT-PCR does not yet show acceptable accuracy. However, RT-PCR tests can take many hours or days to finalise the test outcomes by which time the health condition and infectious status of confirmed patients may deteriorate. Rather than seeking a new single rapid test that improves on RT-PCR, an alternative approach could be to use results from many different profiling tests that are already available and which can be performed quickly by existing equipment [6, 35-36]. How best to use the resulting such multidimensional data is currently controversial.

Rapid blood and serology testing of clinical samples by current equipment enables monitoring of many peripheral blood parameters of interest, some of which indicate changes in organ functions and are used to diagnose a range of conditions diseases [7]. This raises the possibility that such profiling of blood samples could provide predictive information about the disease trajectory and risk of comorbidities for patients with COVID-19. Some data is already used in physician deliberations, but the many available test parameters suggest that an agnostic statistical or machine learning approach would improve the quality of those decisions. We, therefore, undertook a comprehensive assessment that examined the utility of a range of statistical and ML approaches. Indeed, our work identified algorithms that showed significantly improved outcome estimates. This work has, therefore, the potential to optimise the decision processes regarding patient care by clinicians that are under significant time and resource pressure during the current COVID-19 pandemic situation.
    
\section{Materials and Methods}
We used two different datasets in our work, the first of which included data from 89 patients and second from 1,945 patients of confirmed positive COVID-19 cases identified by RT-PCR testing. In the first dataset \cite{b8} we use some statistical methods like student’s t-test, chi-square test and Pearson’s correlation to identify the most significant and associative blood parameters that can strongly distinguish between COVID-19 positive patients and healthy people.  In the second dataset \cite{b9} we use some statistical methods and several machine learning models to identify similar significant blood parameters that can predict severe patients between critical and non-critical COVID-19 positive patients. Figure \ref{fig:workflow} depicts the schematic diagram of the overall workflow of our approach.

We formularized the task of identifying severe COVID-19 patients for their appropriate ward selection as a classification problem by training machine learning models with features from COVID-19 patients’ data. Raw data were collected from the datasets of interest underwent a data wrangling pipeline including denoising, missing value imputation, transformation, normalization, and partition. Next, several statistical comparisons and correlation methods were adopted for feature engineering, including Student's t-test, chi-square test, and Pearson’s correlations. Next, each dataset was further split into three categories based on criteria of existing Non-Communicable Disease (NCD): with NCD, without NCD and all data. Finally, a range of state-of-the-art machine learning methods were trained and evaluated. The algorithms used included decision tree (DT), random forest (RF), gradient boosting machine (GBM), extreme gradient boosting machine (XGBoost), support vector machine (SVM), light gradient boosting machine (LGBM), k-nearest neighbours (KNN), and artificial neural network (ANN)-based deep learning sequential models . Each of these steps are discussed in the following subsections.

\begin{figure*}
  \includegraphics[width=\linewidth]{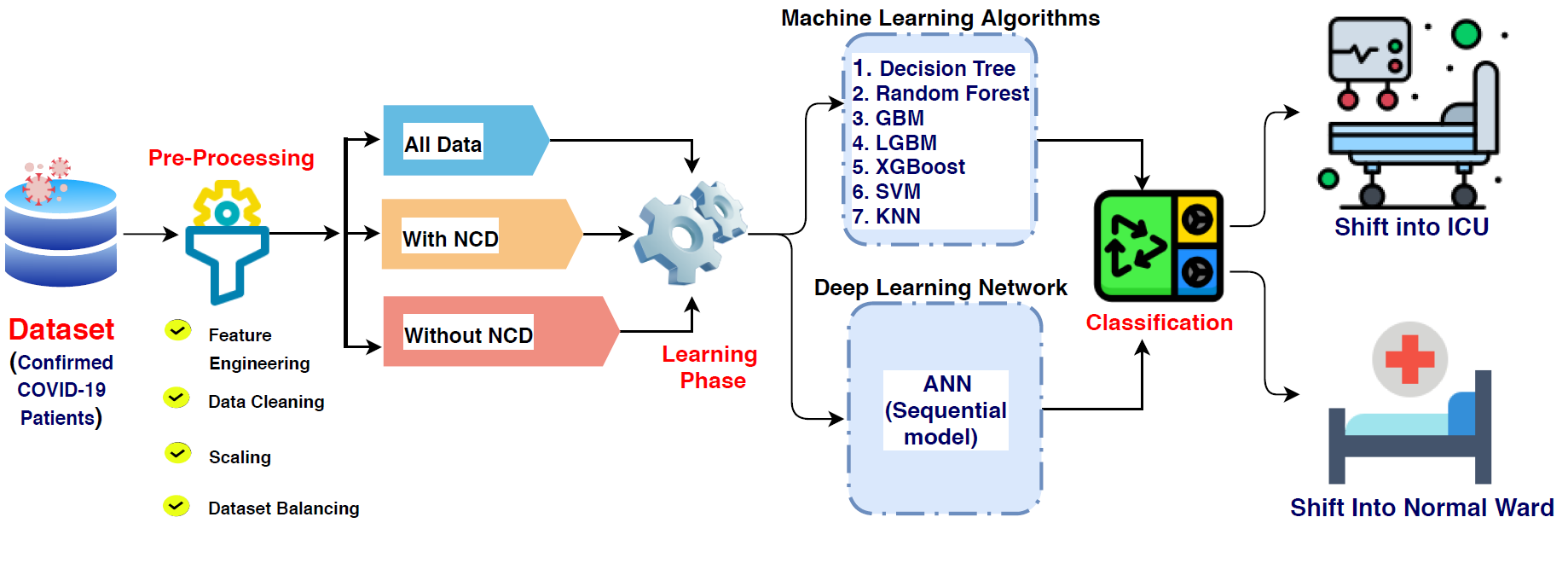}
  \caption{Proposed methodology and workflow of this research for machine learning analysis. [NCD = Non-Communicable Disease]}
  \label{fig:workflow}
\end{figure*}
 
\label{sec:1}

\subsection{Data Collection}
\label{sec:2}
We have collected two different COVID-19 patients’ datasets.  The first was produced by Zenodo \cite{b8}, and contains demographic information and blood sample information from 89 COVID-19 positive patients. In this dataset, 31 patients were alive at the point of data collection, while 58 patients had died prior to collection. The second larger dataset was obtained from the Kaggle online resource \cite{b9}, which contains the grouped information of previous diseases, blood sample results, and vital sign data of 1,945 COVID-19 positive patients. The primary sources of this dataset are Brazilian hospitals that include Sirio Libanes, São Paulo and Brasilia. The parameters of the dataset included patient age percentiles, gender and demographic information. Some patients have pre-existing non-communicable diseases (NCD), including hypertension and immunocompromised status. The blood parameters examined included the following: serum albumin, base excess arterial, base excess venous, bicarbonate arterial, bicarbonate venous, bilirubin, blast, calcium, creatinine, labels of free fatty acids (FFA), Gamma-glutamyl transferase (GGT), glucose, hematocrit, haemoglobin, international normalised ratio (INR), lactate, numbers of leukocytes, lymphocytes neutrophils, partial pressure oxygen (PO2) arterial, PO2 venous, partial pressure carbon dioxide (PCO2) arterial, PCO2 venous,  pH arterial,  pH venous, platelets, potassium, O2 saturation arterial, O2 saturation venous, sodium,  aspartate aminotransferase (AST/TGO), alanine aminotransferase (ALT/ TGP), TTPA, urea, heart rate, respiratory rate, temperature, oxygen saturation, systolic and diastolic blood pressure. During the feature engineering phase in our study, all of these blood parameters were considered as features.

\subsection{Identification of significant routine blood parameters for SARS-CoV-2 infections}
\begin{figure}
  \includegraphics[width=\linewidth]{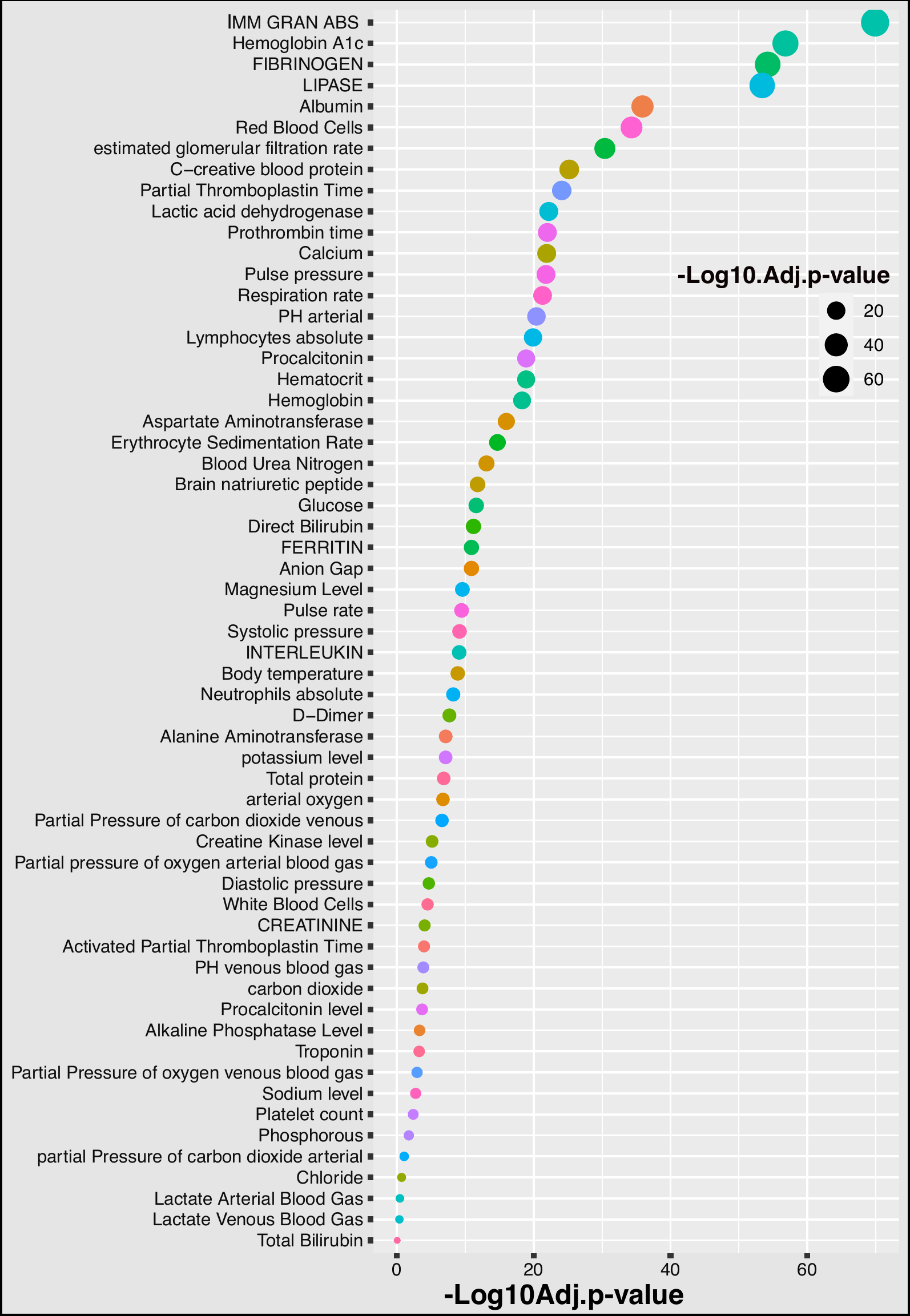}
  \caption{Association of blood parameters to be COVID-19 patients. Parameter measurements and significant difference (using t-test) for different blood parameters between the COVID-19 affected patients compared to the levels expected in regular patients blood parameters.}
  \label{fig:90pValues}
\end{figure}

\subsection{Data Processing}
For the Zenodo dataset \cite{b8} that consists of 89 COVID-19 positive patients, we first removed any unwanted parameters (e.g., ethnicity, BMI, drinking or smoking habit, etc.). We then eliminated all those missing values, resulting in a dataset of 70 patients. In the Sirio Libanes dataset \cite{b9} from Kaggle, there were 1,945 individual patients with 54 types of tests. The primary dataset contained a large number of missing values. The rationale behind the removal of entries with missing parameter value is that, when we conducted a pilot study with the imputation of missing values with mean, median, or regression values, a poor predictive performance was observed. Therefore, we eliminated entries which contained at least one missing value. This elimination resulted in 545 sets of patient data entries in the second dataset that contains no missing values. Among the patients of this dataset, 264 had symptoms severe enough to be admitted into ICU. Both datasets underwent a denoising step, where we removed unwanted strings. Standard scaling techniques performed feature scaling that is the mean value of the data will be 0 with variance value 1; this will be calculated as a mean value of a feature is subtracted from original value then divided by standard deviation. After preprocessing, we consider 545 patients data for the analysis. For the precise study, then we divided this dataset according to whether a patient had a coexisting non-communicable disease (NCD) or not (no NCD). We found 264 NCD patients and 281 without NCDs patients, among which NCD and no NCD groups included 156 and 108 patients (respectively) that were classed as displaying severe conditions. After this data preparation and preprocessing, we have considered all these data for the statistical analysis, and we randomly selected 80\% of the grouped patients data for model training and the rest for model validation testing.

\subsection{Statistical methods to identify the most significant and associative blood parameters}
In the statistical analysis, we used chi-square tests for categorical variables, Student's t-test for continuous variables and Pearson’s correlation among various blood samples counts. The null hypothesis was: the data from both of the COVID-19 patient and healthy population were indistinguishable. Significant blood parameters were chosen based on $P value < 0.05$, while in some cases the selection criteria were adj. $p-value < 0.05$ and absolute value log2 fold-change$(LFC) < 1$. Furthermore, hierarchical clustering was conducted on Pearson’s correlation coefficients for significant grouping parameters \cite{b10,b11,b12}.

\begin{figure*}
  \includegraphics[width=.9\linewidth]{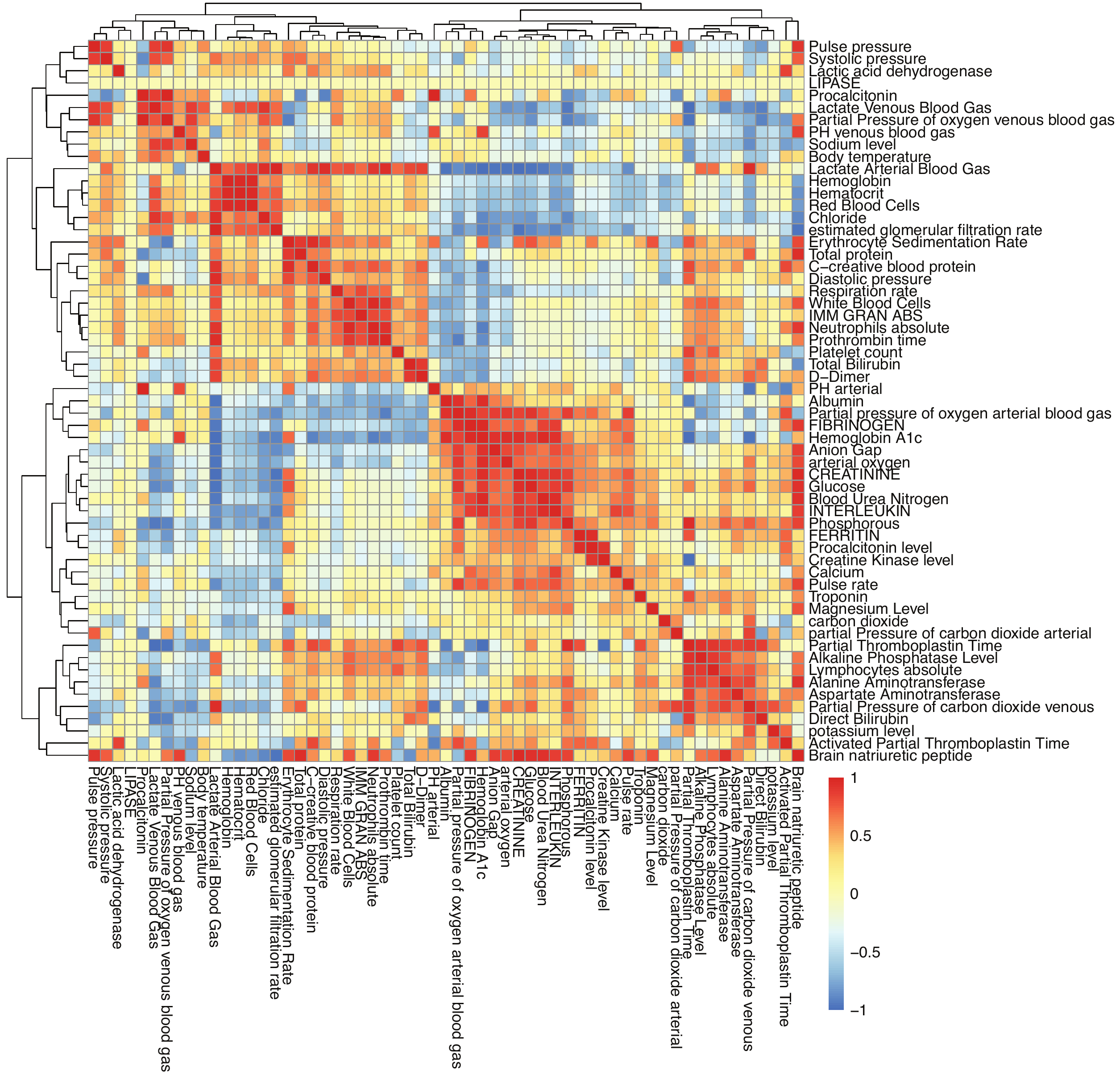}
  \caption{Correlation heat-maps among the various blood parameters examined using the dataset of 89 patients.}
  \label{fig:90pCorr}
\end{figure*}

\subsection{Machine learning models to classify COVID-19 patient severity}
To identify a set of important blood samples as a feature selection step, we have employed a set of machine learning (ML) algorithms using COVID-19 dataset that included data from severe and non-severely affected patients. We chose those ML algorithms that are supervised in performing classification tasks with superior performance and fast execution \cite{b13,b34}. For this purpose, we have considered a basic ensemble learning approach that is based on max voting, averaging, and weighted-averaging and the advanced ensemble learning algorithms works with stacking, blending, bagging, and boosting. The ensemble learning algorithms are the combinations of one or more basic algorithms that are high performing, efficient, effective, and easy to debug \cite{b14,b15}.

We next address the parameters of ML algorithms that were used when it was run. In the DT algorithm, we used the random state as 42, criterion as ‘gini,’ and the minimum sample split was 2. Similarly, in the RF algorithm, the minimum sample split was two, and the number of estimators was 100. The SVM algorithm sets a linear kernel, cache size 200, and 3rd degree. In the GBM algorithm, the learning rate was 0.1, criterion ‘$friedman_mse$,’ and the number of estimators 100. The learning rate in the LGBM algorithm was 0.05, feature fraction 0.9, bagging fraction 0.8, and bagging frequency 5. In the XGB algorithm, we used a tree-based booster, and max depth was six, learning rates 0.1, and the number of estimators 1000.  For the KNN algorithms, the matrices we used were Minkowski; weights were uniform, and the number of neighbours was 3 (k=3).

We have also experimented with a deep learning model, i.e. a feed-forward one-dimensional artificial neural network (ANN), e.g., sequential model. This model consists of an input layer, three hidden layers, and an output layer \cite{b16}. Each layer contains a collection of parallel processing nodes, called neurons, which takes input from the nodes of the previous layer. All the hidden layers are ReLU (Rectified Linear Units)-activated, and the output layer is softmax activated, providing the class probability of the input sample. The network was trained in 1000 epochs using the stochastic gradient descent (SGD) optimization algorithm with categorical cross-entropy loss as a convergence indicator, and a learning rate of 0.0001.

\begin{figure*}
  \includegraphics[width=\linewidth]{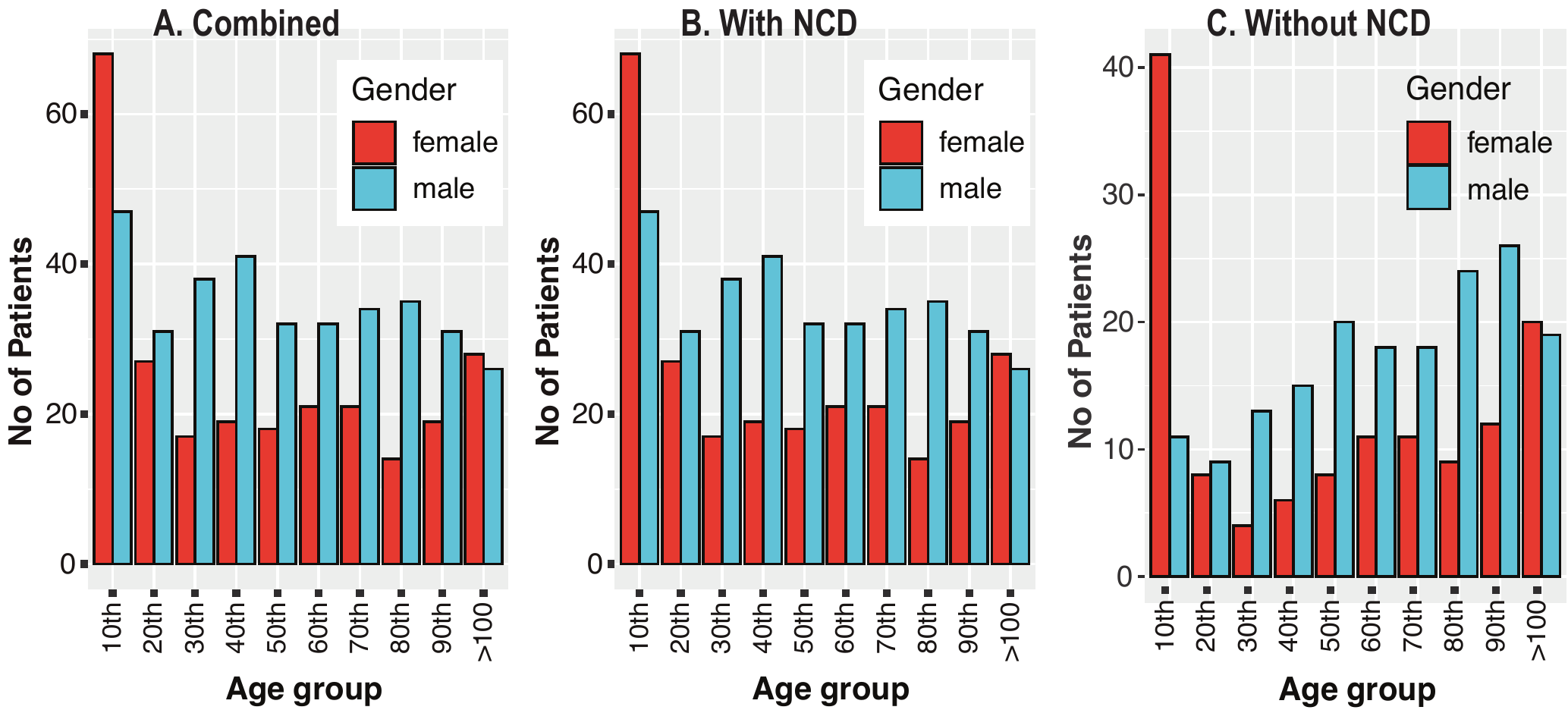}
  \caption{Age distribution of COVID-19 patients for every patients group}
  \label{fig:demog}
\end{figure*}

\subsection{Evaluation matrices for machine learning models}
We evaluated the performance of our models using precision, recall, f1 score, the area under the receiver operator characteristics curve (AUC-ROC) and log-loss function. The precision indicates the proportion of true positive instances among all the positive instances \cite{b17} and the recall indicates the proportion of what the actual true instances measure from the whole instances \cite{b17}. Between precision and recall, we calculated the f1 score metrics for better evaluations. This is the harmonic mean of precision and recall \cite{b17}. The AUC of a classifier is equivalent to the likelihood that the classifier will rank a randomly picked positive value higher than a randomly picked negative value [18]. Logarithmic Loss or log-loss value is used in multinomial logistic regression. The classifier gives probability to each class for all the class samples. If the p is the number of samples which are belonging to q numbers of classes. Then the log-loss function can be calculated as:
\begin{equation}
Log-loss = -\frac{1}{p}\sum_{i=1}^{p}\sum_{j=1}^{q}x_{ij}log(y_{ij})
\end{equation}
where $x_{ij}$ denotes whether p belongs to class i or not, and $y_{ij}$ denotes whether q belongs to class j. Log Loss closer to 0 demonstrates higher accuracy. Without proper evaluation of the ML model utilizing various matrices and relying on the accuracy, it can bring an issue when the model is conveyed on unseen data and can bring about inaccurate predictions.

\section{Results}
We worked with two scenarios of COVID-19 patients analysis, in the first scenario we applied Student’s t-test and Pearson’s correlation to COVID-19 positive patients blood cell parameters and normal range of blood cell parameters we found that both tests support predictive value of C-reactive blood protein, procalcitonin, erythrocyte sedimentation rate, brain natriuretic peptide, ferritin, and D-Dimer are significant for COVID-19 positive patients. In the second scenario, we accounted for only COVID-19 patients for severity calculation. We also applied two different analysis approaches. These were Student’s t-test, the other  a machine learning method  of these approaches found that respiratory rate, lactate, blood pressure (systolic and diastolic), haemoglobin, hematocrit, base excess venous and arterial, neutrophils, albumin, urea, platelet count and potassium were good indicators of patient  disease severity and a small set of a predictor of COVID-19 severity measurements.

The demographic information for the patient data is shown in Table \ref{table:demogTable}, comparing the patients with severe and non-severe symptoms. The patients (n=545) included 198 (36.33\%) were female, 257 (47.16\%) were above 65 years of age, and 264 (48.44\%) were admitted into ICU. Among the group that included only patients with no NCDs (n=281), 107 (38.08\%) were female, and 108 (38.43\%) patients were admitted to ICU. Moreover, in the patient group that had one or more NCDs (n=264), 167 (63.26\%) patients are over 65 years old, and 156 (59.09\%) were admitted into the ICU beds. The age distribution is shown in Figure \ref{fig:demog}.

\begin{figure}
  \includegraphics[width=\linewidth]{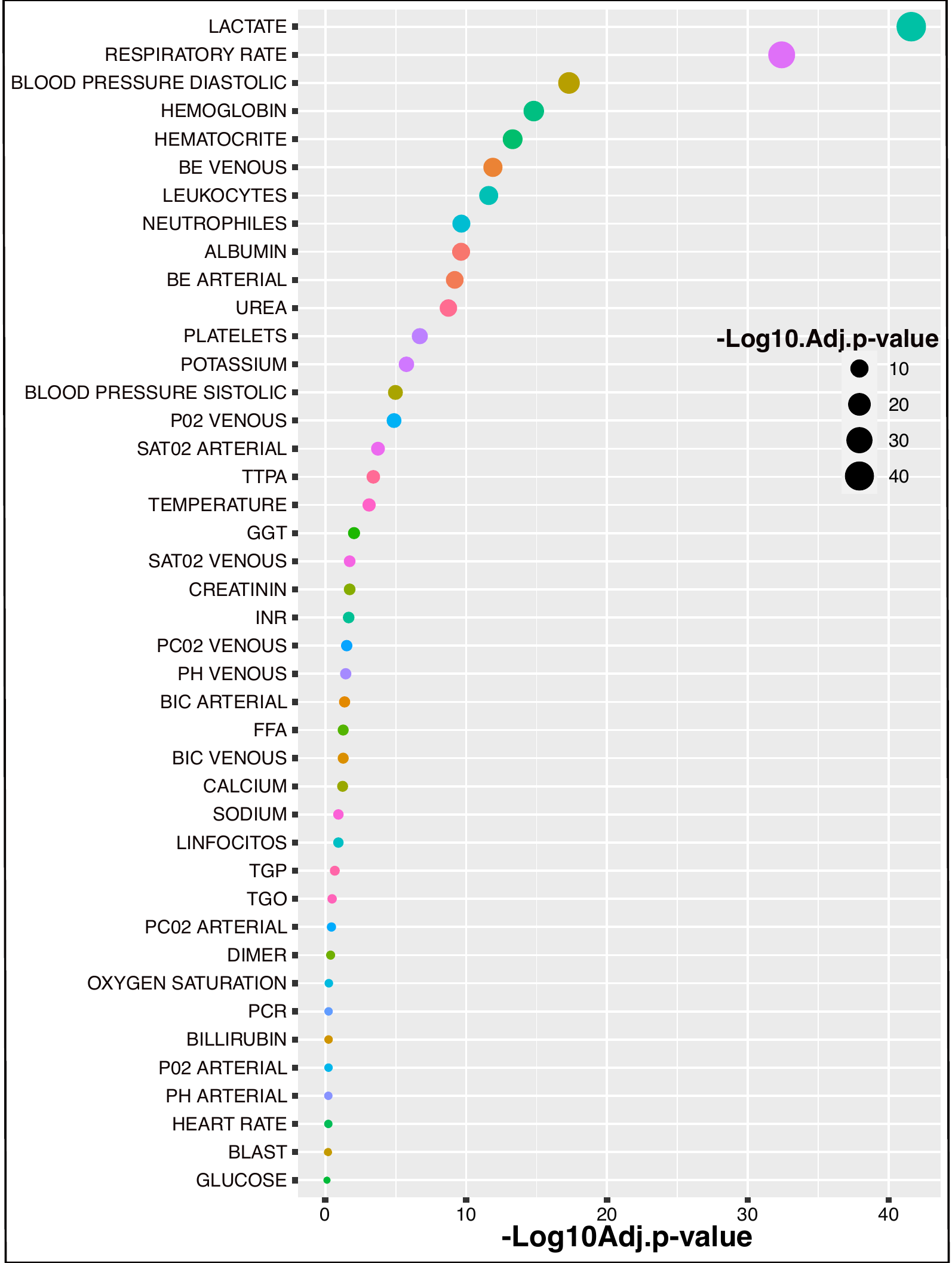}
  \caption{Association of blood parameters with the severity of the COVID-19 disease. Association and significant difference (using t-test) for the different blood parameters between the severe COVID-19 affected patients compared to the non-severe COVID-19 patient blood parameters.}
  \label{fig:severepValues}
\end{figure}

Our first dataset contained 89 confirmed COVID-19 positive patient blood parameters. We hence performed a Student's t-test of tested blood parameters to compare the expected range values (shown in Figure \ref{fig:90pValues}) with COVID-19 patients from the first dataset. According to the t-test analyses, the more likely informative blood parameters were immature grans (abs), haemoglobin A1C, fibrinogen, lipase, albumin, red blood cell count, estimated glomerular filtration rate, C-reactive protein and partial thromboplastin time. The combination of t-test and log2 fold changes (LFC) indicated that the eight (08) most significant candidate predictive parameters for COVID-19 positive status were lipase, C-reactive protein, procalcitonin level, erythrocyte sedimentation rate, brain natriuretic peptide, ferritin, D-Dimer, and creatine kinase level,  which all showed p-values less than 0.001 and absolute $LFC>1$.

\begin{table*}[h!]
\centering
\begin{tabular}{|c c c c|} 
 \hline
 \myrowcolour
 & All patients (545) & Without NCD (281) & With NCD (264) \\ [0.5ex]
 \myrowcolour
 & n(\%) & n(\%) & n(\%) \\
 \hline
Age above 65 years & 257(47.16) & 90(32.03) & 167(63.26)\\
 \myrowcolour
Age Percentile &  &  & \\
10 & 115(21.1) & 63(22.42) & 52(19.7)\\
 \myrowcolour
20 & 58(10.64) & 41(14.59) & 17(6.44)\\
30 & 55(10.09) & 38(13.52) & 17(6.44)\\
 \myrowcolour
40 & 60(11.01) & 39(13.88) & 21(7.96)\\
50 & 50(9.17) & 22(7.83) & 28(10.61)\\
 \myrowcolour
60 & 53(9.73) & 24(8.54) & 29(10.98)\\
70 & 55(10.09) & 26(9.25) & 29(10.98)\\
 \myrowcolour
80 & 49(8.99) & 16(5.7) & 33(12.5)\\
90 & 50(9.17) & 12(4.27) & 38(14.39)\\
 \myrowcolour
$\geq$100 & 54(9.91) & 15(5.34) & 39(14.77)\\
Gender &  &  & \\
 \myrowcolour
Female & 198(36.33) & 107(38.08) & 91(34.47)\\
Male & 347(63.67) & 174(61.92) & 173(65.53)\\
 \myrowcolour
Admitted into ICU & 264(48.44) & 108(38.4) & 156(59.09)\\
 \hline
\end{tabular}
\caption{Accuracy and evaluation matrices for every patients groups}
\label{table:demogTable}
\end{table*}
We applied Student's t-test to the a second dataset to attempt to discriminate severe and non-severe symptom COVID-19 positive patients by finding  patients characteristics which are associated with the target variable disease severity;  the analysis results are shown in Figure \ref{fig:severepValues}. According to t-test results from the most significant features of COVID-19 patients that imply the patient severity is the patient's hospital visit identifier, age, gender, previous co-existing diseases with hypertension. The most significant blood parameters according to t-test results were lactate, respiratory rate, diastolic blood pressure, haemoglobin, hematocrit, base excess venous, leukocytes, neutrophils, albumin, base excess arterial, urea, platelets count, potassium and systolic blood pressure.

\subsection{Clustering and co-expression analysis}
We also performed Pearson’s correlation between the different routine blood parameters. The Pearson correlation is shown in Figure \ref{fig:90pCorr}. And we found that there are some hierarchical clusters of the tests that showed significance equally for all the patients. From the total 59 blood samples, we found that there were four (04) different most concordant clusters that are strongly correlated with each other. The first cluster comprised   pulse pressure and systolic blood pressure. The second comprised haemoglobin, hematocrit, and red blood cells. The third cluster was built with C-reactive protein, erythrocyte sedimentation rate, diastolic blood pressure, and respiratory rate. Procalcitonin levels, ferritin, and creatine kinase levels composed the fourth cluster.

\subsection{Prediction of severe COVID-19 patients for critical treatment using machine models}
In this section, we first describe the performances of used machine learning models and their applications. We then present the most important reduced set of blood and physical sign parameters that can precisely discriminate the patients with severe disease from those with non-severe disease. The reduced collection of blood parameters, also known as significance for severe COVID-19 positive patients’.

\begin{table*}[h!]
\centering
\begin{tabular}{|c c c c c c c c c c|} 
 \hline
  \myrowcolour
 Dataset & Matrices & RF & LGBM & SVM & DT & XGB & GBM & KNN & ANN \\ [0.5ex] 
 \hline
 Combined & Accuracy & 0.89 & 0.88 & 0.84 & 0.82 & 0.88 & 0.89 & 0.84 & 0.83\\
  \myrowcolour
 & AUC & 0.89 & 0.88 & 0.84 & 0.82 & 0.88 & 0.89 & 0.84 & 0.82\\
 & Precision & 0.9 & 0.88 & 0.84 & 0.83 & 0.91 & 0.91 & 0.81 & 0.92\\
  \myrowcolour
 & Recall & 0.9 & 0.9 & 0.88 & 0.83 & 0.86 & 0.88 & 0.93 & 0.69\\
 & F1-Score & 0.9 & 0.89 & 0.86 & 0.83 & 0.88 & 0.89 & 0.86 & 0.79\\
  \myrowcolour
 & Log Loss & 3.8 & 4.12 & 5.39 & 6.34 & 4.12 & 3.8 & 5.39 & 6.02\\
 \hline
With NCDs & Accuracy & 0.93 & 0.91 & 0.84 & 0.86 & 0.91 & 0.88 & 0.74 & 0.74\\
 \myrowcolour
 & AUC & 0.92 & 0.91 & 0.83 & 0.85 & 0.9 & 0.86 & 0.73 & 0.71\\
 & Precision & 0.89 & 0.91 & 0.83 & 0.85 & 0.89 & 0.84 & 0.74 & 0.86\\
  \myrowcolour
 & Recall & 1 & 0.94 & 0.91 & 0.91 & 0.97 & 0.97 & 0.81 & 0.48\\
 & F1-Score & 0.94 & 0.92 & 0.87 & 0.88 & 0.93 & 0.9 & 0.78 & 0.62\\
  \myrowcolour
 & Log Loss & 2.42 & 3.02 & 5.45 & 4.85 & 3.03 & 4.24 & 9.09 & 9.09\\
 \hline
Without NCDs & Accuracy & 0.91 & 0.93 & 0.84 & 0.84 & 0.87 & 0.89 & 0.77 & 0.74\\
 \myrowcolour
 & AUC & 0.91 & 0.92 & 0.83 & 0.84 & 0.87 & 0.89 & 0.79 & 0.71\\
 & Precision & 0.89 & 0.89 & 0.83 & 0.85 & 0.82 & 0.82 & 0.65 & 0.77\\
  \myrowcolour
 & Recall & 0.97 & 1 & 0.91 & 0.88 & 0.85 & 0.9 & 0.85 & 0.82\\
 & F1-Score & 0.93 & 0.94 & 0.87 & 0.86 & 0.83 & 0.86 & 0.74 & 0.79\\
  \myrowcolour
  & Log Loss & 3.03 & 2.42 & 5.45 & 5.45 & 4.56 & 3.91 & 7.82 & 9.12\\
 \hline
\end{tabular}
\caption{Accuracy and evaluation matrices for every data groups}
\label{table:accuracyTable}
\end{table*}

For the machine learning analysis for the second dataset, we applied the respective methods and models; their performances and the evaluation matrices are shown in Table \ref{table:accuracyTable}. In the data group of all patients with and without NCDs, we found that RF and GBM gave the highest testing accuracy score 89\% and the other methods and models perform with above 80\% testing accuracy score. The highest Area Under the ROC curve (AUC) score is for RF, and GBM 89\% and also other methods and models scored ethical AUC values above 80\%. The highest precision values were seen for XGB and GBM with 91\%. The highest Recall values are for KNN 93\%, RF and LGBM score 90\% and other methods scores above 80\%. The best F1-score is 90\% for RF and others have above 80\%. RF and GBM have the lowest log loss value of 3.8\%, and other methods and models also have ethical lowest values below 7\%. In this patient group, we saw that all of our applied models achieve good performance in every evaluation matrix with above 80\% accuracy score, so in practice any of those can be employed.

In the data group of patients who have no NCDs, we found that RF performed with the highest accuracy score of 93\%, LGBM and XGB performed with 91\% and SVM, DT performed with a good accuracy score above 80\%. Nevertheless, KNN and ANN showed comparatively low accuracy scores 74\%, because when we divide the dataset the size of data being small. RF performed with the highest AUC score, 92\%, LGBM 91\% and XGB 90\%. LGBM have the highest precision value, 91\%, RF and XGB have 89\%. The highest precision values were91\% for LGBM, and other methods and models have above 80\% except for KNN (74\%). The highest recall values were for RF 100\%, XGB and GBM 97\%, and other methods and models had above 80\%, except ANN (48\%). RF scores highest F1-score 94\%, XGB 93\%, LGBM 92\% and SVM and DT scores 88\%. However, KNN and ANN scored comparatively lowly, with 78\% and 62\% respectively because of lower training sample sizes. The lowest log loss value was 2.42\% for RF, and other methods and models also perform with good log loss values below 10\%. In this patient group, we observed that excepting KNN and ANN all of the model accuracy scores were above 80\%, and the evaluation matrix showed good model performances. So, the best performing models could usefully be applied in clinical scenarios.

In the data group of patients who had one or more coexisting NCDs, we found that LGBM performed with highest accuracy score 93\%, and RF, GBM, XGB, SVM, and DT performed with 91\%, 89\%, 87\%, 84\%, and 84\%, respectively. KNN and ANN performed poorly but this was due to the small amount of available data showing 77\% and 74\% accuracy, respectively. The highest AUC score was 92\% for LGBM and RF, SVM, DT, XGB, GBM, KNN and ANN scored 91\%, 83\%, 84\%, 87\%, 89\%, 79\% and 71\% respectively. RF and LGBM perform with highest precision value 89\% and other methods and models performed with good precision values above 80\% except for KNN and ANN. LGBM scores highest recall value 100\%, RF 97\%, GBM 90\%, SVM 83\%, and DT 88\% and the other methods and models perform above 80\%. The highest F1-score is 94\% for LGBM, and RF also performs with 93\% and other methods and models perform above 80\% except for KNN and ANN. KNN and ANN performed with 74\% and 79\% F1-score, respectively, but the number of training samples were small.

\begin{figure*}
  \includegraphics[width=\linewidth]{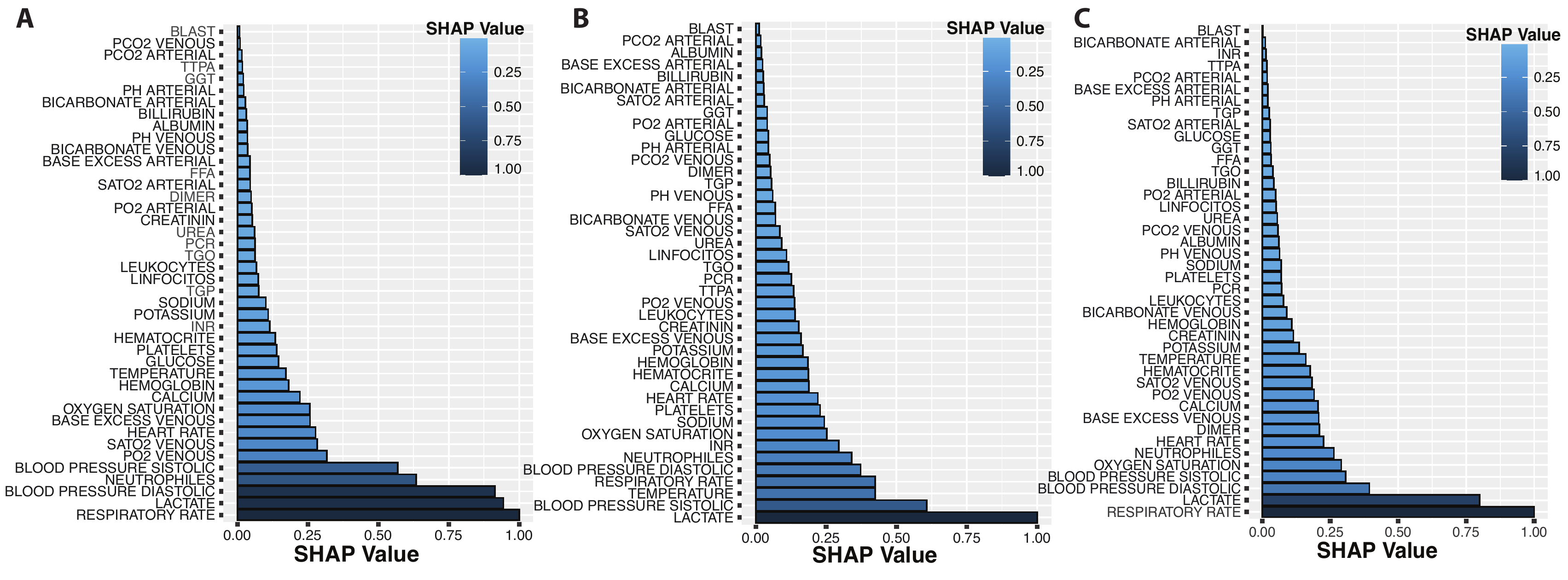}
  \caption{Sorted significant and impacted blood parameters COVID-19 patients. Artificial intelligence models predicted the blood parameters most predictive for the severity of COVID-19 symptoms. The higher coefficient values of ML model outcomes indicate the higher significant association of seriousness. A. Combined patients group; B. Patients those have NCDs; C. Patients with no NCDs; Note, the SHAP values are defined as the coefficient values of each parameter after model training.}
  \label{fig:shapAll}
\end{figure*}

In Figure \ref{fig:shapAll}. A, we present the most crucial blood parameters data that are highly predictive for COVID-19 patient disease severity among patients with and without NCDs. For the combined (i.e., both with and without NCD) patient data group, we found that RF method found significant features were respiratory rate, lactate, diastolic blood parameters, neutrophiles, base excess venous, systolic blood pressure, temperature, platelets, venous $O_{2}$ saturation, venous $PO_{2}$ saturation, calcium, haemoglobin, heart rate, potassium, and TGO . LGBM found lactate, diastolic blood pressure, neutrophils, systolic blood pressure, PO2 venous, calcium, base excess venous, hematocrit, respiratory rate, potassium, heart rate, platelets, bilirubin, sodium, and TGP as significant. Using SVM, the significant blood parameters  were diastolic blood pressure, heart rate, respiratory rate, systolic blood pressure, oxygen saturation level, neutrophiles, base excess venous, saturation $O_{2}$ venous, leukocytes, lactate, $PO_{2}$ venous, bicarbonate venous, base excess arterial, hematocrit and PH arterial. The most predictive blood test data for the DT algorithm were respiratory rate, neutrophiles, glucose, INR, haemoglobin, temperature, and TGP. For the XGB algorithm, significant blood parameters were diastolic blood pressure, lactate, neutrophiles, $PO_{2}$ venous, calcium, systolic blood pressure, platelets, hematocrit, respiratory rate, PCR, saturation $O_{2}$ venous, urea, TGO, base excess venous, and lymphocytes. The most significant blood parameters for the GBM algorithm were respiratory rate, diastolic blood pressure, systolic blood pressure, lactate, neutrophiles, calcium, $PO_{2}$ venous, oxygen saturation level, platelets, base excess venous, heart rate, lymphocytosis, TGO, haemoglobin and sodium. KNN identified, lactate, respiratory rate, oxygen saturation level, systolic blood pressure, saturation $O_{2}$ venous, diastolic blood pressure, $PO_{2}$ venous, base excess venous, haemoglobin, neutrophils, lymphocytosis, PH venous, saturation $O_{2}$ arterial, $PO_{2}$ arterial, and glucose as significant. The most striking predictive blood parameters for ANN were lactate, respiratory rate, saturation $O_{2}$ venous, diastolic blood pressure, heart rate, systolic blood pressure, potassium, sodium, hematocrit, base excess arterial, creatinine, haemoglobin, albumin and, $PO_{2}$ arterial.

In the data group of the patients who have one or more coexisting NCDs, the most significant blood parameters are shown in Figure \ref{fig:shapAll}. B. For this data group, the Rf method found respiratory rate, lactate, base excess venous, neutrophils, diastolic blood pressure, calcium, leukocytes, temperature, $PO_{2}$ venous, sodium, potassium, bicarbonate venous, systolic blood pressure, urea and hematocrit as significant blood parameters. The LGBM method found lactate, neutrophils, haemoglobin, creatinine, FFA, platelets, temperature, leukocytes, calcium, systolic blood pressure, TGO, INR, hematocrit, potassium and diastolic blood pressure as significant. The most highly predictive blood parameters when using the SVM method was systolic blood pressure, heart rate, oxygen saturation level, respiratory rate, diastolic blood pressure, lymphocytosis, base excess venous, bicarbonate venous, hematocrit, sodium, saturation O2 venous, bicarbonate arterial, temperature, PH venous and base excess arterial. The DT method finds INR, systolic blood pressure, and temperature were significant. The significant blood parameters for the XGB method were lactate, neutrophils, diastolic blood pressure, systolic blood pressure, PCR, TGO, calcium, TTPA, creatinine, sodium, lymphocytosis, respiratory rate, platelets, potassium and urea. The GBM method found lactate, respiratory rate, diastolic blood pressure, systolic blood pressure, neutrophiles, oxygen saturation level, TTPA, heart rate, urea, calcium, $PO_{2}$ venous, dimer, sodium, hematocrit, and TGO as significant. KNN identified temperature, oxygen saturation level, hematocrit, sodium, diastolic blood pressure, haemoglobin, potassium, systolic blood pressure, platelets, lactate, heart rate, lymphocytosis, TTPA, calcium and creatinine as significant. The significant blood parameters for the ANN model were lactate, platelets, systolic blood pressure, sodium, potassium, $PO_{2}$ venous, diastolic blood pressure, PH arterial, leukocytes, saturation $O_{2}$ venous, heart rate, $PCO_{2}$ venous, $PO_{2}$ arterial, creatinine and neutrophils.

In the data group of patients who have no NCDs, the most significant blood parameters are shown in Figure \ref{fig:shapAll}. C and described in this section. We found that the RF method identified respiratory rate, lactate, base excess venous,$PO_{2}$ venous, saturation $O_{2}$ venous, calcium, diastolic blood pressure, $PCO_{2}$ venous, bicarbonate venous, haemoglobin, systolic blood pressure, potassium, $PO_{2}$ arterial, hematocrit and leukocytes as significant. The LGBM method found lactate, respiratory rate, systolic blood pressure, $PO_{2}$ venous, calcium, diastolic blood pressure, hematocrit,
haemoglobin, neutrophils, potassium, base excess venous, urea, oxygen saturation level,  creatinine and leukocytes were significant. The most important blood parameters for the SVM method were respiratory rate, diastolic blood pressure diastolic, oxygen saturation level, systolic blood pressure, base excess venous, temperature, lactate, haemoglobin, neutrophils, bicarbonate venous, sodium, glucose, hematocrit and creatinine. DT found dimer, neutrophiles and oxygen saturation level as significant components. The most significant blood parameters for XGB method is a respiratory rate, lactate, hematocrit, calcium, systolic blood pressure, creatinine, potassium, PCR, diastolic blood pressure, neutrophils, haemoglobin, temperature, PH venous, venous$O_{2}$ saturation  and TGO. The GBM method found respiratory rate, diastolic blood pressure, systolic blood pressure, lactate, calcium, $PO_{2}$ venous, oxygen saturation level, base excess venous, hematocrit, potassium, PH venous,$PCO_{2}$ venous, leukocytes, PCR and TGO were significant. The KNN method identified respiratory rate, lactate, oxygen saturation level, heart rate, venous $O_{2}$ saturation creatinine, neutrophiles, bicarbonate venous, lymphocytosis, temperature, sodium, albumin, bilirubin, and FFA as significant. The most significant blood parameters for ANN were lactate, heart rate, respiratory rate, saturation $O_{2}$ venous, platelets, albumin, potassium, base excess venous, diastolic blood pressure, GGT, dimer, FFA, saturation $O_{2}$ and leukocytes arterial.

We also observed that beyond blood parameters, some vital signs of COVID-19 patients also predicted disease severity. Patient gender, age, coexisting diseases (such as hypertension) were also found to be significant for these patient data groups of data.

\section{Discussion}

After the worldwide outbreak of COVID-19, classification of disease mortality risk are of very great significance in prevention, and treatment allocation. In this investigation, we found a number of blood analysis parameters that can be used as risk factors for the assessment of likely disease severity in COVID-19 patients. We developed predictive algorithms that utilised a large number of blood parameters and demonstrated that these methods and the potential to predict the disease severity of COVID-19 patients with high accuracy.

We identified a number of features of patient data that contributed strongly to the predicted value of the algorithms (i.e., were found to contribute to the accuracy of all our best machine learning algorithms) some of whom were not obvious candidate predictors.  We found that the absolute value of lymphocytes in the severe symptom group of patients was consistently lower than that of the non-severe symptom group. The neutrophil parameters of the severe symptom group was higher than that of the non-severe symptom group. The latter characteristics indicated a level of immune activation heightened and possibly playing a role in the “inflammatory storm” that is a characteristic of COVID-19 severe symptoms which result in very high tissue and cell harm \cite{b19}. Low lymphocyte levels may reflect impeded antibody-based immune cell functions    suspected to result in patients with severe COVID-19 susceptible to bacterial infection \cite{b20}. The algorithm outcomes indicated that numbers of circulating lymphocytes in the patients who developed severe symptoms was significantly lower than those who did not have severe symptoms. In contrast, the inclusion of neutrophils in the severe ICU patients' was influenced higher, which is consistent with the consequences of Qin et al. \cite{b21}.

We found that the indicator factors could be reliable to predictors that discriminated between severe and non-severe COVID-19 patients. Recent work has uncovered the utility of routine blood parameters in the screening of COVID-19 patients. This is facilitated by the fact that blood parameter analysis is generally fast, affordable, and promptly accessible in the same health facility where patients are receiving treatment. The pathological tests of COVID-19 patients identified some abnormalities in some blood parameters. Previous published data has identified a number of altered blood parameters inCOVID-19 patients who develop severe symptoms include in addition to the lymphocyte and neutrophil parameters noted above, are, eosinophils, basophils, monocytes, platelets and total leukocytes as well as serum levels of urea, potassium, haemoglobin and C-reactive blood protein. \cite{b22,b23,b24}, which provides supportive evidence for our findings. Li et al. identified that COVID-19 affects bacterial pneumonia in some cases of mortality [25]. Bacterial contamination also causes expanded leucocyte count and neutrophil count, which may be linked to defective immune responses. A few patients with COVID-19 have abnormal blood coagulation function: prothrombin time is increased, and D-dimer is raised [19]; thrombosis is linked with expanded platelet consumption and diminished platelet number.

Respiratory rate is one of the principal vital signs for patient symptom severity in COVID-19. Abnormally high respiratory rates (<12 or >25 breaths/min) are also seen in a range of conditions including  asthma, heightened anxiety, pneumonia, congestive heart failure, lung disease (all of which exacerbate COVID-19 conditions when presenting as comorbidities)  and are a significant feature  in severely affected COVID-19 patients \cite{b26,b27}. Elevated heart rate is similarly a key sign [28] and may be the cause of   dizziness or shortness of breath in COVID-19 patients [29]. Blood pressure is additionally a clinical sign for COVID-19 patients \cite{b30}. Hypoxemia is also a sign that indicates the average below level of oxygen saturation in the blood. The usual range of arterial oxygen is 75 to 100 (mm Hg) approximately, and in pulse oximeter reading the expected range from 95\% to 100\%; below 90\% indicates patients' condition is critical \cite{b31}, and this is often observed in COVID-19 patients who may lack other obvious symptoms, making it a particularly dangerous feature of the disease. Related to this is the serum lactic acid test, which is also a significant test that indicates disease severity in COVID-19 patients. Typically the level of lactate in the blood is very low, but when it rises it is typically associated with low oxygen levels \cite{b32,b33}.

In summary there are a number of signs and symptoms that can indicate COVID-19 is likely to become severe in a patient. What is needed is a standardised and objective way to combine these and other less obvious predictors in a way that can optimise patient outcomes and resource management. Our methodology, described here and derived from a number of different machine learning algorithms, can provide such an improved method. Indeed the fact that high accuracy was obtained using similar predictors by different machine learning algorithms (so there is limited sensitivity to methodology) can give confidence that these parameters are of utility, and the approach is a sound one.

\vspace{-0.5cm}
\section{Conclusion}
The results of our analysis indicated that there is a strong relationship between particular abnormal blood parameters and disease severity status in hospitalised patients with COVID-19. The primary utility of our findings is that the subset of routine blood parameters linked to disease severity could be used in a predictive algorithm that would better enable appropriate care to be given before the onset of severe symptoms. This is of particular importance in developing countries, where the hospitals where ICU beds are a limited resource. This can be achieved using a relatively small number of currently available hospital blood-based tests to proper utilisation of ICU resources and identifying patients that need to be monitored closely.

Among the association between blood parameters that are able to give predictive information regarding COVID-19 symptoms severity, the levels of C-reactive blood protein appeared to have the strongest predictive value. Levels of haemoglobin, procalcitonin, erythrocyte sedimentation rates, brain natriuretic peptide, ferritin, D-Dimer, and platelets were likewise showed significant deviation from the normal control group in this. Other parameters, respiratory rate, lactate, blood pressure (systolic and diastolic), hematocrit, base excess venous and arterial, neutrophils, albumin, and urea showed less obvious deviations but clearly had predictive value. Our work suggests that links between these parameters and COVID-19, and perhaps similar pro-inflammatory infectious diseases are worth more detailed physiological investigations. 

There were a few limitations to our study. Firstly, the small sample size may restrict the precision of the severity identification. Secondly, the absence of more detailed clinical information in the datasets that were used (such as patients’ age, sex and comorbidities), which may hinder better classification, although it suggests that future studies with new datasets could address this and improve on our work. Finally, the disease severity and mortality of COVID-19 varies significantly from country to country, for reasons that are very poorly understood but suggests that this type of predictive analysis should be carried out on data from other parts of the world to improve performance of the algorithm. Nevertheless, we hope our study can be used by practitioners, facilitates sound policy-makers to improve resource allocation and patient outcomes for COVID-19 patients.

\bibliographystyle{vancouver}
\bibliography{\jobname}

\begin{thebibliography}{}

\bibitem{b1} Mohammadi M, Meskini M, Pinto ALdo N. 2019 Novel coronavirus (COVID-19) overview. Journal of Public Health. doi:10.1007/s10389-020-01258-3
\bibitem{b2} Yang J, Chen X, Deng X, et al. Disease burden and clinical severity of the first pandemic wave of COVID-19 in Wuhan, China. Nature Communications. October 2020. doi:10.1038/s41467-020-19238-2
\bibitem{b3} Ahamad MM, Aktar S, Rashed-Al-Mahfuz M, et al. A machine learning model to identify early stage symptoms of SARS-Cov-2 infected patients. Expert Systems with Applications. 2020;160:113661. doi:10.1016/j.eswa.2020.113661
\bibitem{b4} Prin M, Wunsch H. International comparisons of intensive care: informing outcomes and improving standards. Current Opinion in Critical Care. 2012;18(6):700-706. $doi:10.1097/MCC.0b013e32835914d5$
\bibitem{b5} Hong KH, Lee SW, Kim TS, et al. Guidelines for Laboratory Diagnosis of Coronavirus Disease 2019 (COVID-19) in Korea. Annals of Laboratory Medicine. 2020;40(5):351-360. doi:10.3343/alm.2020.40.5.351
\bibitem{b6} Butt C, Gill J, Chun D, Babu BA. RETRACTED ARTICLE: Deep learning system to screen coronavirus disease 2019 pneumonia. Applied Intelligence. 2020. doi:10.1007/s10489-020-01714-3
\bibitem{b7} Li Z, Yi Y, Luo X, et al. Development and clinical application of a rapid IgM‐IgG combined antibody test for SARS‐CoV‐2 infection diagnosis. Journal of Medical Virology. 2020;92(9):1518-1524. doi:10.1002/jmv.25727
\bibitem{b8} Stachel A. Development and validation of a machine learning model for use as an automated artificial intelligence tool to predict mortality risk in patients with COVID-19. Zenodo. http://doi.org/10.5281/zenodo.3893846. Published June 14, 2020. Accessed November 16, 2020.
\bibitem{b9} Sírio-Libanês H. COVID-19 - Clinical Data to assess diagnosis. Kaggle. https://www.kaggle.com/S\%C3\%ADrio-Libanes/covid19. Published June 22, 2020. Accessed November 16, 2020.
\bibitem{b10} Nihan ST. Karl Pearsons chi-square tests. Educational Research and Reviews. 2020;15(9):575-580. $doi:10.5897/err2019.3817$
\bibitem{b11} Horne AD. Statistics, Use in Immunology. Encyclopedia of Immunology. 1998:2211-2215. $doi:10.1006/rwei.1999.0559$
\bibitem{b12} 11. Correlation and regression. The BMJ. https://www.bmj.com/about-bmj/resources-readers/publications/statistics-square-one/11-correlation-and-regression. Accessed November 16, 2020.
\bibitem{b13} Patel HH, Prajapati P. Study and Analysis of Decision Tree Based Classification Algorithms. International Journal of Computer Sciences and Engineering. 2018;6(10):74-78. $doi:10.26438/ijcse/v6i10.7478$
\bibitem{b14} Aluja-Banet T, Nafria E. Stability and scalability in decision trees. Computational Statistics. 2003;18(3-4):505-520. $doi:10.1007/bf03354613$
\bibitem{b15} Sciabola S, Fang C. Gradient boosting decision tree models for better temporal ADME prediction from an industrial perspective. 2020. doi:10.1021/scimeetings.0c06777
\bibitem{b16} Hutter F, Hoos HH, Leyton-Brown K. Sequential Model-Based Optimization for General Algorithm Configuration. Lecture Notes in Computer Science Learning and Intelligent Optimization. 2011:507-523. $doi:10.1007/978-3-642-25566-3_40$
\bibitem{b17} Wang R, Li J. Bayes Test of Precision, Recall, and F1 Measure for Comparison of Two Natural Language Processing Models. Proceedings of the 57th Annual Meeting of the Association for Computational Linguistics. 2019. $doi:10.18653/v1/p19-1405$
\bibitem{b18} Verbakel JY, Steyerberg EW, Uno H, et al. ROC curves for clinical prediction models part 1. ROC plots showed no added value above the AUC when evaluating the performance of clinical prediction models. Journal of Clinical Epidemiology. 2020;126:207-216. doi:10.1016/j.jclinepi.2020.01.028
\bibitem{b19} Mo P, Xing Y, Xiao Y, et al. Clinical characteristics of refractory COVID-19 pneumonia in Wuhan, China. Clinical Infectious Diseases. 2020. doi:10.1093/cid/ciaa270
\bibitem{b20} Chen N, Zhou M, Dong X, et al. Epidemiological and clinical characteristics of 99 cases of 2019 novel coronavirus pneumonia in Wuhan, China: a descriptive study. The Lancet. 2020;395(10223):507-513. doi:10.1016/s0140-6736(20)30211-7
\bibitem{b21} Qin C, Zhou L, Hu Z, et al. Dysregulation of Immune Response in Patients with COVID-19 in Wuhan, China. SSRN Electronic Journal. 2020. doi:10.2139/ssrn.3541136
\bibitem{b22} Aljame M, Ahmad I, Imtiaz A, Mohammed A. Ensemble learning model for diagnosing COVID-19 from routine blood tests. Informatics in Medicine Unlocked. 2020;21:100449. doi:10.1016/j.imu.2020.100449 
\bibitem{b23} Li X, Wang L, Yan S, et al. Clinical characteristics of 25 death cases with COVID-19: A retrospective review of medical records in a single medical center, Wuhan, China. International Journal of Infectious Diseases. 2020;94:128-132. doi:10.1016/j.ijid.2020.03.053
\bibitem{b24} Sun S, Cai X, Wang H, et al. Abnormalities of peripheral blood system in patients with COVID-19 in Wenzhou, China. Clinica Chimica Acta. 2020;507:174-180. doi:10.1016/j.cca.2020.04.024
\balance
\bibitem{b25} Li X, Wang L, Yan S, et al. Clinical characteristics of 25 death cases with COVID-19: A retrospective review of medical records in a single medical center, Wuhan, China. International Journal of Infectious Diseases. 2020;94:128-132. doi:10.1016/j.ijid.2020.03.053
\bibitem{b26} Bernardi L, Porta C, Gabutti A, Spicuzza L, Sleight P. Modulatory effects of respiration. Autonomic Neuroscience. 2001;90(1-2):47-56. doi:10.1016/s1566-0702(01)00267-3
\bibitem{b27} Lee M. Clinical Characteristics Of Early Noncritical Hospitalized Patients With Coronavirus Disease 2019 (Covid-19): A Single-Center Retrospective Study In New York City. 2020. doi:10.26226/morressier.5ebc261fffea6f735881a237
\bibitem{b28} Peer N, Lombard C, Steyn K, Levitt N. Elevated resting heart rate is associated with several cardiovascular disease risk factors in urban-dwelling black South Africans. Scientific Reports. 2020;10(1). doi:10.1038/s41598-020-61502-4
\bibitem{b29} Pavri BB, Kloo J, Farzad D, Riley JM. Behavior of the PR interval with increasing heart rate in patients with COVID-19. Heart Rhythm. 2020;17(9):1434-1438. doi:10.1016/j.hrthm.2020.06.009
\bibitem{b30} Lazic S, Lazic B. The correlation between systolic and diastolic blood pressure and diastolic parameters in arterial hypertension in the presence of normal systolic function. Cardiologia Croatica. 2014;9(5-6):166-166. doi:10.15836/ccar.2014.166
\bibitem{b31} Anusha B (2017) Assessment of Pulp Oxygen Saturation Levels by Pulse Oximetry for Pulpal Diseases –A Diagnostic Study. Journal Of Clinical And Diagnostic Research. doi: 10.7860/jcdr/2017/28322.10572
\bibitem{b32} Aktar S, Ahamad MM, Talukder A, et al. Machine Learning and Meta-Analysis Approach to Identify Patient Comorbidities and Symptoms that Increased Risk of Mortality in COVID-19. arXiv. 2020; arXiv:2008.12683
\bibitem{b33} Tan L, Kang X, Ji X, et al. Validation of Predictors of Disease Severity and Outcomes in COVID-19 Patients: A Descriptive and Retrospective Study. Med. 2020. doi:10.1016/j.medj.2020.05.002
\bibitem{b34} Uddin S, Khan A, Hossain ME, Moni MA. Comparing different supervised machine learning algorithms for disease prediction. BMC Medical Informatics and Decision Making. 2019;19(281). doi:10.1186/s12911-019-1004-8
\bibitem{b35} Nain Z, Rana HK, Liò P, Islam SMS, Summers MA, Moni MA. Pathogenetic profiling of COVID-19 and SARS-like viruses. Briefings in Bioinformatics. August 2020. doi:10.1093/bib/bbaa173
\bibitem{b36} Taz TA, Ahmed K, Paul BK, et al. Network-based identification genetic effect of SARS-CoV-2 infections to Idiopathic pulmonary fibrosis (IPF) patients. Briefings in Bioinformatics. 2020. doi:10.1093/bib/bbaa235

\end{thebibliography}

\end{document}